# Bounds for Approximation in Total Variation Distance by Quantum Circuits

E. Knill, knill@lanl.gov*

July 1995

**Abstract**

It was recently shown that for reasonable notions of approximation of states and functions by quantum circuits, almost all states and functions are exponentially hard to approximate [5]. The bounds obtained are asymptotically tight except for the one based on total variation distance (TVD) . TVD is the most relevant metric for the performance of a quantum circuit. In this paper we obtain asymptotically tight bounds for TVD. We show that in a natural sense, almost all states are hard to approximate to within a TVD of $2/e-\epsilon$ even for exponentially small $\epsilon$. The quantity $2/e$ is asymptotically the average distance to the uniform distribution. Almost all states with probability amplitudes concentrated in a small fraction of the space are hard to approximate to within a TVD of $2-\epsilon$. These results imply that non-uniform quantum circuit complexity is non-trivial in any reasonable model. They also reinforce the notion that the relative information distance between states (which is based on the difficulty of transforming one state to another) fully reflects the dimensionality of the space of qubits, not the number of qubits.

## 1 Introduction

Given two probability distributions $\nu$ and $\mu$ on a finite event space, the total varation distance (TVD) between $\nu$ and $\mu$ is defined by $|\nu-\mu|_1 = \sum_x |\nu(x) -$

*Address: MS B265, Los Alamos National Laboratory, Los Alamos, NM 87545. This work was performed under the auspices of the U.S. Department of Energy under Contract No. W-7405-ENG-36.

$\mu(x)|$. The TVD on an event space with $n$ elements is equivalent to the $L_1$ metric on

$$\Delta(n) = \{\mathbf{x} \in \mathbf{R}^n \mid \mathbf{x} \geq \mathbf{0},\ \mathbf{x} \cdot \mathbf{1} = 1\},$$

where $\mathbf{R}$ is the set of real numbers, the expression $\mathbf{x} \geq \mathbf{0}$ means that each coordinate $x_i$ of $\mathbf{x}$ satisfies $x_i \geq 0$, $\mathbf{y} \cdot \mathbf{z}$ is the inner product of $\mathbf{y}$ and $\mathbf{z}$, and $\mathbf{0}$ ($\mathbf{1}$) is the vector with all entries 0 (1).

Our interest in the TVD comes from the theory of quantum circuits. The domain $H_n$ of computation of a quantum circuit is an $n$-fold tensor product of *qubits* $Q$, $H_n = Q^{\otimes n}$. A qubit $Q$ is the two dimensional complex Hilbert space generated by the basis vectors $|0\rangle$ and $|1\rangle$. The standard basis of $H_n$ consists of elements of the form $|b_1\rangle|b_2\rangle \ldots |b_n\rangle$ which we abbreviate $|b_1 b_2 \ldots b_n\rangle$ or $|b\rangle$ if $b \in 2^{[n]}$ is an $n$-bit vector or a number written in binary (in reverse order). A quantum circuit applies a unitary operation to $H_n$ by composing a number of primitive unitary operations called *quantum gates*. A quantum gate with $g$ inputs is a unitary operator $V$ on $H_g$. The specification of the circuit describes which $g$ qubits each gate should act on. The gate's action is obtained by identifying $H_n$ with $H_g \otimes H_{n-g}$, where $H_g$ is the factor corresponding to the $g$ input qubits. The gate acts as $V \otimes \mathbf{I}$ on $H_g \otimes H_{n-g}$, where $\mathbf{I}$ is the identity matrix. The complexity of the quantum circuit is the number of gates applied. An important property is that all unitary operations are exactly representable as compositions of 2-qubit gates [4]. See [8, 7, 2] for more detailed descriptions and motivations.

The output of a quantum circuit is a state in $H_n$. The knowledge that can be gained from a state is restricted to what can be learned by *measuring* it. A measurement on the first $m$ bits of a state $|x\rangle$ induces a probability distribution on $m$-bit vectors defined by

$$\mathrm{Prob}_m(b \mid |x\rangle) = \sum_{b'} |\langle bb'|x\rangle|^2$$

for $b \in 2^{[m]}$. The goal of a computation is to transform an input state $|b\rangle$ to some output state $|x(b)\rangle$ whose induced distribution $\mathrm{Prob}_m(\cdot \mid |x(b)\rangle)$ is sufficiently close to a desired one. Since we are comparing probability distributions, the TVD is the most appropriate distance measure to use for evaluating the success of the computation.

It is shown in [5] that even if $|0\rangle$ is the only input of interest, almost no distribution on $2^{[m]}$ can be approximated within a TVD of $\frac{1}{2} - \epsilon$ unless the number of gates in the circuit is exponentially large in $m$. The notion of "almost no distribution" is derived from the induced Lebesgue measure on $\Delta(2^m)$. This result is suboptimal in two ways. First one can compute the minimum expected distance of $x \in \Delta(n)$ from a fixed point in $\Delta(n)$ as $\frac{2}{e} - o(1) > \frac{1}{2}$. Thus the situation where a small number of gates can approximate a distribution to better than average is not excluded. Second, most computationally interesting distributions are highly concentrated. Such distributions are on average within a TVD $2 - o(1)$ of other distributions. Finding an approximation within distance 1 (say) might already be good. For approximating functions with large domains, the results for classical approximation problems in [5] show that even weak approximation is difficult. However, for small domains, the worst-case complexity of approximating highly concentrated distributions to within $2 - \epsilon$ total variation distance was left open. In this paper we resolve both of these issues by showing that the number of gates must be nearly exponential for any non-trivial approximation to be achieved for a non-negligible fraction of possible input-output relationships.

Our proofs are based on the same arguments as those given in [5], and use lemmas given there. The new results in this paper are obtained by making use of a large deviation argument to show that random elements of $\Delta(n)$ have certain properties with respect to the TVD.

## 2 Main results

We begin with some definitions.

For $N' < N$, there are $\binom{N}{N'}$ ways of embedding $\Delta(N')$ in $\Delta(N)$. For an $N'$ tuple $S \subseteq [n]$, let $\Delta(S, N)$ be the face of $\Delta(N)$ consisting of the vectors $\mathbf{x}$ which satisfy that $x_i > 0$ iff $i \in S$. Let $\Delta(N', N) = \bigcup_{S:|S|=N'} \Delta(S, N)$. Note that $\Delta(N, N) = \Delta(N)$ and for $N' \leq N''$, $\Delta(N', N) \subseteq \Delta(N'', N)$.

Let $\Delta(N', N)^k$ be the set of $k$-tuples of members of $\Delta(N', N)$. We endow $\Delta(N', N)^k$ with the measure $\mu$ obtained by normalizing the Lebesgue measure so that $\mu(\Delta(N', N)^k) = 1$. This is the natural uniform distribution on $\Delta(N', N)^k$. The Lebesgue measure is denoted by $\nu$. In general we will use $\mu$ to denote the uniform distribution and $\nu$ the induced Lebesgue measure on the polytope of interest. If necessary, we will subscript $\mu$ by the polytope being considered. When using probabilistic concepts defined on a polytope, we always mean the uniform distribution.

We extend the TVD to $\Delta(N)^k$ by

$$|\mathbf{x}-\mathbf{y}|_1 = \frac{1}{k}\sum |\mathbf{x}_i - \mathbf{y}_i|_1,$$

which is the average TVD of the components. Here $\mathbf{x}_i$ denotes the $i$'th member of the $k$-tuple of elements $\mathbf{x}$ of $\Delta(N)^k$.

A unitary operator $U$ acting on $H_n$ induces a map which takes the first $k$ basis elements $|0\rangle, \ldots, |k-1\rangle$ to the $k$-tuple of probability distributions $\mathrm{Prob}_m(\cdot \mid U|0\rangle), \ldots, \mathrm{Prob}_m(\cdot \mid U|k-1\rangle)$ in $\Delta(2^m)^k$. Denote this $k$-tuple by $\phi_m(U)$.

Let $G_g^{(b,n)}$ be the set of unitary operators on $n$ qubits expressible as a composition of at most $b$ $g$-input quantum gates. Let $X(b,g,m,n;N,d,k)$ consist of the members of $\Delta(N,2^m)^k$ which are within average TVD $d$ of an element of $\phi_m(G_g^{(b,n)})$. The number of inputs $g$ is assumed to be constant in the discussions below.

Note that for the purpose of bounding $X(b,g,m,n;N,d,k)$ from above we can assume that $b \geq (n-m)/g$. Otherwise some input qubits which do not participate in the final measurement are involved in the computation and may be eliminated. To avoid other trivial cases we assume that $b \geq n > m \geq 1$, $N \geq 2$ and $k \geq 1$.

**Theorem 2.1** *There exist constants $c_i > 0$ such that for $0 < \epsilon \leq 2$,*

$$\ln(\mu_{\Delta(2^m)^k}(X(b,g,m,n;2^m,2/e-\epsilon,k))) \leq 2^{c_1 g} b \ln(2b/\epsilon) + c_2 mk - c_3\epsilon^2 2^m k.$$

Lemma 3.1 shows that $2/e - o(1)$ is the average distance of $\mathbf{y} \in \Delta(2^m)$ to the uniform distribution $\mathbf{1}/2^m$.

The proof of the theorem can be used to find explicit values of the constants[1]. We do not make any attempts to optimize the inequalities in this paper.

**Corollary 2.2** *For $0 < \alpha < 1$, almost all $k$-tuples of states require $2^{\alpha m(1-o(1))}$ $g$-input gates for approximation by a quantum circuit on the first $m$ qubits to within a TVD of $2/e - 2^{-(1-\alpha)m/2}$.*

**Theorem 2.3** *There exist constants $c_i > 0$ such that for $N = \gamma 2^m$ and $0 < \epsilon \leq 2$,*

$$\begin{aligned} &\ln(\mu_{\Delta(N,2^m)^k}(X(b,g,m,n;N,2-\epsilon,k))) \\ &\qquad \leq \; 2^{c_1 g} b \ln(2b/\epsilon) + c_2 mk - ((c_3\epsilon - c_4\gamma^{1/4})\gamma 2^m k. \end{aligned}$$

[1] These values turn out not to be excessively large or small.

**Corollary 2.4** *Let $3/4 < \alpha < 1$ and $\gamma = o(2^{-4(1-\alpha)m})$. Consider those $k$-tuples of states $|x\rangle$ which satisfy that $\text{Prob}_m(\cdot \mid |x\rangle)$ has at most $\gamma 2^m$ non-zero values. Then almost all such $k$-tuples of states require at least $\gamma 2^{\alpha m(1-o(1))}k$ $g$-input gates for approximation by a quantum circuit on the first $m$ qubits to within a TVD of $2 - 2^{-(1-\alpha)m}$.*

**Proofs of Theorems 2.1 and 2.3.** The proofs of the theorems closely follow those given for Theorems 4.4 and 4.5 in [5]. We outline of the proofs, deferring the proofs of the lemmas to the following section.

First note that if we represent a unitary operator by the composition of $b$ fixed gates, we have at most $\binom{n}{g}^b$ choices for ways of composing them. This gives a bound on the number of structurally distinct quantum circuits. The next observation is that the group of unitary operators on $g$ qubits can be densely covered using a constant (for fixed $g$) number of operators. This is formalized by Lemma 4.4 of [5] which we state next. For any linear operator $U$, let $||U||_2$ denote the two-norm of $U$ defined by

$$||U||_2 = \max_{\mathbf{x}:|\mathbf{x}|=1} |U\mathbf{x}|.$$

**Lemma 2.5** *There exists a subset $U_{g,\delta}$ of $G_g$ with no more than $(2/\delta)^{2^{4g}}$ elements such that for every $V \in G_g$ there exists a $U \in U_{g,\delta}$ satisfying that $||U - V||_2 \le \delta$.*

The lemma's relevance to the problem at hand is due to the relationship between the two-norm and the TVD, and the behavior of the two-norm under composition of unitary operators. The two-norm satisfies

$$|\text{Prob}(\cdot \mid U|b\rangle) - \text{Prob}(\cdot \mid V|b\rangle)|_1 \le 2||U - V||_2$$

(Lemma 2.2 of [5]) and for unitary operators $U_i$ and $V_i$

$$||U_1U_2 - V_1V_2||_2 \le ||U_1 - V_1||_2 + ||U_2 - V_2||_2$$

(Lemma 2.3 of [5]).

Let $d = 2/e - \epsilon$ for Theorem 2.1 and $d = 2 - \epsilon$ for Theorem 2.3. Let $B_{\mathbf{x}}(d) = \{\mathbf{y} \mid |\mathbf{x} - \mathbf{y}|_1 < d\}$. Let $X = X(b, g, m, n; N, d, k)$. Then $X$ is included in the union of the balls $B_{\phi_m(U)}((1+\alpha)d)$, where $U$ ranges over the unitary operators defined by those circuits of at most $b$ elements for which each gate is in $U_{g,\alpha d/(2b)}$.

Choose $\alpha = \epsilon/(2d)$. First consider the statement of Theorem 2.1. By Theorem 3.11 there are constants $c_i > 0$ such that

$$\ln(\mu_{\Delta(N,2^m)^k} B_{\phi_m(U)}((1+\alpha)d) \cap \Delta(N, 2^m)^k)) \leq -(c_1\epsilon^2 2^m - c_2 m)k,$$

which implies

$$\begin{aligned}
\mu(X) &\leq \binom{n}{g}^b (4b/(\alpha d))^{2^{4g}b}\, e^{-(c_1\epsilon^2 2^m - c_2 m)k}, \\
\ln(\mu(X)) &\leq b(\ln(n) + 2^{4g}\ln(4b/(\alpha d))) - c_1\epsilon^2 2^m k + c_2 mk \\
&\leq 2^{c_3 g} b \ln(2b/\epsilon) - c_1\epsilon^2 2^m k + c_2 mk.
\end{aligned}$$

This proves Theorem 2.1.

To prove Theorem 2.3, we can proceed in a similar fashion. Let $\gamma = N/2^m$. By Theorem 3.12 there are constants $c_1$, $c_2$ and $c_3$ such that for sufficiently large $m$,

$$\ln\mu(B_{\phi_m(U)}((1+\alpha)d) \cap \Delta(N, 2^m)^k) \leq -(c_1\epsilon - c_2\gamma^{1/4})\gamma 2^m k + c_3 mk.$$

Hence

$$\begin{aligned}
\mu(X) &\leq \binom{n}{g}^b (4b/(\alpha d))^{2^{4g}b}\, e^{-(c_1\epsilon - c_2\gamma^{1/4})\gamma 2^m k + c_3 mk} \\
\ln(\mu(X)) &\leq 2^{c_3 g} b \ln(2b/\epsilon) + c_3 mk - (c_1\epsilon - c_2\gamma^{1/3})\gamma 2^m k.
\end{aligned}$$

■

## 3 Large Deviation Bounds For Total Variation Distance

For the remainder of the paper we assume that $N \geq 2$.

### 3.1 The Expectation of $|\mathbf{x} - \mathbf{y}|_1$

For fixed $\mathbf{x}$, let

$$D(\mathbf{x}, N', N) = \int d\mu_{\Delta(N',N)}(\mathbf{y})|\mathbf{x} - \mathbf{y}|_1.$$

be the expected TVD of $\mathbf{x}$ from elements of $\Delta(N', N)$. Write $D(\mathbf{x}, N) = D(\mathbf{x}, N, N)$.

**Lemma 3.1** *For* $\mathbf{x} \geq \mathbf{0}$, $D(\mathbf{x}, N) = \sum_i \frac{2}{N}(1 - x_i)^N + \sum_i x_i - 1$.

**Proof.** We have $|\mathbf{x}-\mathbf{y}|_1 = \sum_i |x_i - y_i|$, so by additivity of expectations, we can consider each coordinate separately. The induced density function of the distribution of $y_i$ is $(N-1)(1-t)^{N-2}$. The contribution of the $i$'th coordinate to $D(\mathbf{x}, N)$ is

$$\begin{aligned}&\int_0^1 dt(N-1)|t-x_i|(1-t)^{N-2}\\ &\quad = (N-1)\int_0^{x_i} dt(x_i-t)(1-t)^{N-2} + (N-1)\int_{x_i}^1 dt(t-x_i)(1-t)^{N-2}.\end{aligned}$$

We have

$$(N-1)\int dt(x_i-t)(1-t)^{N-2} = (\tfrac{1}{N}(1-t)-(x_i-t))(1-t)^{N-1} + C,$$

so that the contribution of the first coordinate is

$$\mathrm{Exp}(|x_i-y_i|) = \frac{2}{N}(1-x_i)^N - (\frac{1}{N}-x_i).$$

■

**Corollary 3.2** $D(\mathbf{1}/N, N) = 2/e - O(1/N)$.

**Lemma 3.3** *For* $\mathbf{x} \geq \mathbf{0}$, $D(\mathbf{x}, N', N) = \sum_i \frac{2}{N}(1-x_i)^{N'} + \sum_i x_i - 1$.

**Proof.** This is a direct application of Lemma 3.1. The contribution of $x_i$ is

$$\begin{aligned}&\frac{1}{\binom{N}{N'}}\left(\sum_{S:i\in S}\mathrm{Exp}(|x_i-y_i| : \mathbf{y}\in\Delta(S,N)) + \sum_{S:i\notin S} x_i\right)\\ &\quad = \frac{\binom{N-1}{N'-1}}{\binom{N}{N'}}\left(\frac{2}{N'}(1-x_i)^{N'} + x_i - \frac{1}{N'}\right) + \frac{\binom{N-1}{N'}}{\binom{N}{N'}}x_i\\ &\quad = \frac{2}{N}(1-x_i)^{N'} + x_i - \frac{1}{N}.\end{aligned}$$

■

**Lemma 3.4** *For* $\mathbf{x} \in \Delta(N)$, $D(\mathbf{x}, N', N)$ *is minimized by* $\mathbf{x} = \mathbf{1}/N$.

**Proof.** Note that $D(\mathbf{x}, N', N)$ is convex in $\mathbf{x}$. By symmetry, the minimum must be achieved by $\mathbf{x} = \mathbf{1}/N$. ■

### 3.2 On the Distribution of $|\mathbf{x}-\mathbf{y}| : \mathbf{y} \in \Delta(N)$

Let

$$\begin{aligned} T(\mathbf{x}, d, N', N) &= \mu_{\Delta(N',N)}(\mathbf{y} \mid |\mathbf{x}-\mathbf{y}|_1 < d), \\ T(d, N', N) &= T(\mathbf{1}/N, d, N', N), \\ T(\mathbf{x}, d, N) &= T(\mathbf{x}, d, N, N), \\ T(d, N) &= T(\mathbf{1}/N, d, N). \end{aligned}$$

We would like to obtain good upper bounds on $T(\mathbf{x}, d, N)$ for $\mathbf{x} \in \Delta(N)$ and $d = 2/e - \epsilon$.

**Theorem 3.5** *There exist constants $c_i > 0$ such that for any $\mathbf{x} \in \Delta(N)$ $T(\mathbf{x}, 2/e - \epsilon, N) \leq e^{-c_1 \epsilon^2 N + c_2 \ln(N)}$.*

The proof of the theorem requires several lemmas. First we simplify the problem to the case of $\mathbf{x} = \mathbf{1}/N$.

**Lemma 3.6** *For $\mathbf{x}$ such that $\mathbf{x} \cdot \mathbf{1} = \alpha N$, $T(\mathbf{x}, d, N)$ is maximized by $\mathbf{x} = \alpha \mathbf{1}$.*

**Proof.** The proof is by induction on $N$. For $N = 2$, the result follows by inspection. Let $N > 2$. Let $\mathbf{x} = (x_1, \mathbf{x}')$ with $\mathbf{x}' \in \mathbf{R}^{N-1}$. We have

$$\begin{aligned} & \mu(\mathbf{y} \in \Delta(N) \mid |\mathbf{x}-\mathbf{y}|_1 < d) \\ = \; & \mu((y_1, \mathbf{y}') \in \Delta(N) \mid |x_1 - y_1| + |\mathbf{x}' - \mathbf{y}'| < d) \\ = \; & \int_0^1 dt (N-1)(1-t)^{N-2} \\ & \qquad \mu_{(1-t)\Delta(N-1)}(\mathbf{y}' \in (1-t)\Delta(N-1) \mid |\mathbf{x}' - \mathbf{y}'| < d - |x_1 - t|), \end{aligned}$$

where $(1-t)\Delta(N-1) = \{\mathbf{y} \geq \mathbf{0} \mid \mathbf{y} \cdot \mathbf{1} = (1-t)\}$. In the last step we used the fact that the distribution of $y_1$ has density $(N-1)(1-t)^{N-2}$. By induction and scaling, the integrand is maximized by $\mathbf{x}' = (\mathbf{x}' \cdot \mathbf{1})\mathbf{1}/(N-1)$ independently of $y_1$ and $t$. Note that replacing $\mathbf{x}'$ by $(\mathbf{x}' \cdot \mathbf{1})\mathbf{1}/(N-1)$ does not change $\mathbf{x} \cdot \mathbf{1}$. This implies that the probability of interest is maximized if every subset of $N-1$ coordinates of $\mathbf{x}$ is uniform, which is satisfied only by $\mathbf{x} = \alpha \mathbf{1}$. ■

To obtain a bound on $T(d, N)$ requires decomposing $\Delta(N)$ according to which orthant $\mathbf{y} - \mathbf{1}/N$ belongs to. Formally, let $\Delta_k(N)$ be the set of $\mathbf{y} \in \Delta(N)$ such that exactly the first $k$ coordinates of $\mathbf{y} - \mathbf{1}/N$ are positive. Then $\Delta(N)$ is a disjoint union of coordinate permuted copies of the $\Delta_k(N)$. In particular

$$1 = \mu_{\Delta(N)}(\Delta(N)) = \sum_k \binom{N}{k} \mu_{\Delta(N)} \Delta_k(N).$$

We will make use of the tail probabilities of the sums of $N-1$ independent identically distributed uniform random variables. We define them here in the language of polytopes. Let $[0,1] = \{x \in \mathbf{R} \mid 0 \leq x \leq 1\}$ and

$$U(M,s) = \mu_{[0,1]^M}\{\mathbf{x} \in [0,1]^M \ \mid \ \mathbf{x} \cdot \mathbf{1} < s\}.$$

**Lemma 3.7** *There exist constants $a_i > 0$ and $0 < \kappa_0 < 1$ such that for $|k/N - \kappa_0| > \epsilon$,*

$$\binom{N}{k} \mu_{\Delta(N)}(\Delta_k(N)) \leq e^{-a_1 \epsilon^2 N + a_2 \ln(N)}.$$

**Proof.** Note that $\Delta_0(N)$ has measure zero, so we can assume that $k > 0$. Let $X_k = \Delta_k(N) - \mathbf{1}/N$. We project $X_k$ onto the last $N-1$ coordinates and consider its measure in the set $S = \{\mathbf{z} \mid \mathbf{z} \geq -\mathbf{1}/N,\ \mathbf{z} \cdot \mathbf{1} \leq 0\}$. The volume of this set is $1/(N-1)!$. The projection of an element of $X_k$ can be written as $(\mathbf{y}, \mathbf{z})$ with $\mathbf{y} \in \mathbf{R}^{k-1}$ and $\mathbf{z} \in \mathbf{R}^{N-k}$ corresponding to the positive and negative coordinates, respectively.

$$\begin{aligned}
\mu_{\Delta(N)}(\Delta_k(N)) &= \mu_S(X_k) \\
&= (N-1)!\, \nu((\mathbf{y}, \mathbf{z}) \ \mid \ \mathbf{y} \in \mathbf{R}^{k-1},\ \mathbf{y} \geq 0, \\
&\qquad \mathbf{z} \in \mathbf{R}^{N-k},\ \mathbf{1}/N \geq \mathbf{z} \geq 0,\ \mathbf{z} \cdot \mathbf{1} \geq \mathbf{y} \cdot \mathbf{1}) \\
&= (N-1)!/N^{N-1}\, \nu((\mathbf{y}, \mathbf{z}) \ \mid \ \mathbf{y} \in \mathbf{R}^{k-1},\ \mathbf{y} \geq 0, \\
&\qquad \mathbf{z} \in [0,1]^{N-k},\ \mathbf{z} \cdot \mathbf{1} \geq \mathbf{y} \cdot \mathbf{1}) \\
&= (N-1)!/N^{N-1}\, \nu((\mathbf{y}, \mathbf{z}) \ \mid \ \mathbf{y} \in \mathbf{R}^{k-1},\ \mathbf{y} \geq 0, \\
&\qquad \mathbf{z} \in [0,1]^{N-k},\ N-k \geq (\mathbf{y}, \mathbf{z}) \cdot \mathbf{1}),
\end{aligned}$$

where we scaled by $N$ in the second step and obtained the last identity by replacing $\mathbf{z}$ with $\mathbf{z} - \mathbf{1}$. The volume in the last expression can be decomposed according to which translate of the standard hypercube $\mathbf{y}$ is in. We label these translates by the coordinates of the corner nearest the origin and note that by symmetry, only the sum of these coordinates is relevant. This gives

$$\begin{aligned}
\mu_{\Delta(N)}(\Delta_k(N)) &= (N-1)!/N^{N-1} \sum_{l=0}^{N-k} \binom{k+l-2}{k-2} \\
&\qquad \nu(\mathbf{y} \in [0,1]^{N-1} | N-k-l \geq \mathbf{y}\cdot\mathbf{1}) \\
&= (N-1)!/N^{N-1} \sum_{l=0}^{N-k} \binom{k+l-2}{k-2} U(N-1, N-k-l).
\end{aligned}$$

Let

$$C(k,l) = (N-1)!/N^{N-1} \binom{N}{k} \binom{k+l-2}{k-2} U(N-1, N-k-l).$$

We next show that there exist $b_i > 0$, $0 < \kappa_0 < 1$ and $0 < \xi_0 < 1$ such that for $(|k/N - \kappa_0|^2 + |l/N - \xi_0|^2)^{1/2} > \epsilon$, $C(k,l) \leq e^{-b_1\epsilon^2 N + b_2 \ln(N)}$. By summing over $l$, this implies that for $|k/N - \kappa_0| > \epsilon$

$$\binom{N}{k} \mu(\Delta_k(N)) \leq N e^{-b_1\epsilon^2 N + b_2\ln(N)} = e^{-b_1\epsilon^2 N + (b_2+1)\ln(N)},$$

which gives the lemma.

To prove the desired property of $C(k,l)$, consider the functions

$$\begin{aligned}
f(\kappa,\xi,N) &= \ln(C(\lfloor \kappa N \rfloor, \lfloor \xi N \rfloor))/N, \\
f(\kappa,\xi) &= \lim_N f(\kappa,\xi,N),
\end{aligned}$$

with domain $0 \leq \kappa \leq 1$ and $0 \leq \kappa + \xi \leq 1$. Since the sum of the $C(k,l)$ is 1, $f(\kappa,\xi,N) \leq 0$. Let $H_e(x) = -x\ln(x) - (1-x)\ln(1-x)$ be the information function base $e$. Then for some constant $b_2$,

$$f(\kappa,\xi,N) \leq -1 + H_e(\kappa) + (\kappa+\xi)H_e(\kappa/(\kappa+\xi)) + \ln(U(N-1, N(1-\kappa-\xi)))/N + b_2\ln(N)/N,$$

where we applied Lemma A.3 and Stirling's approximation. The term $b_2\ln(N)/N$ accounts for the polynomial factors in Stirling's approximation of $(N-1)!$ as well as the correction for integer rounding in Lemma A.3. By Theorem A.7, $r(x) = -\lim_n \ln(U(n,x))/n$ is convex (where it is finite) and identically 0 for $x \geq 1/2$. In addition $\ln(U(n,xn))/n \leq -r(x)$. Hence

$$\begin{aligned}
f(\kappa,\xi,N) &\leq f_u(\kappa,\xi,N) =_{\text{def}} b_2\ln(N)/N - 1 + H_e(\kappa) \\
&\qquad + (\kappa+\xi)H_e(\kappa/(\kappa+\xi)) - (N-1)/N\; r((1-\kappa-\xi)(N/(N-1))), \\
f(\kappa,\xi) &= -1 + H_e(\kappa) + (\kappa+\xi)H_e(\kappa/(\kappa+\xi)) - r(1-\kappa-\xi).
\end{aligned}$$

To show that $f(\kappa,\xi)$ is strictly concave, we evaluate the Hessian of $g(\kappa,\xi) = f(\kappa,\xi) + r(1-\kappa-\xi)$. Note that $(\kappa+\xi)H_e(\kappa/(\kappa+\xi)) = -\kappa\ln(\kappa) - \xi\ln(\xi) + (\kappa+\xi)\ln(\kappa+\xi)$.

$$
\begin{aligned}
\partial_\kappa g(\kappa,\xi) &= \ln(1-\kappa) - 2\ln(\kappa) + \ln(\kappa+\xi) \\
\partial_\xi g(\kappa,\xi) &= -\ln(\xi) + \ln(\kappa+\xi) \\
\partial_\kappa^2 g(\kappa,\xi) &= -\frac{1}{1-\kappa} - \frac{2}{\kappa} + \frac{1}{\kappa+\xi} \\
&= -\frac{1}{\kappa(1-\kappa)} - \frac{\xi}{\kappa(\kappa+\xi)} \\
\partial_\xi^2 g(\kappa,\xi) &= \frac{-\kappa}{\xi(\kappa+\xi)} \\
\partial_\xi\partial_\kappa g(\kappa,\xi) &= \frac{1}{\kappa+\xi}.
\end{aligned}
$$

The Hessian of $g$ is therefore given by

$$
\begin{bmatrix} -\frac{1}{\kappa(1-\kappa)} - \frac{\xi}{\kappa(\kappa+\xi)} & \frac{1}{\kappa+\xi} \\ \frac{1}{\kappa+\xi} & \frac{-\kappa}{\xi(\kappa+\xi)} \end{bmatrix}.
$$

Thus the diagonal elements of the Hessian are strictly negative for $0 < \xi < \infty$ and $0 < \kappa < 1$. Its determinant is given by

$$
\tfrac{-\kappa}{\xi(\kappa+\xi)}\left(-\tfrac{1}{\kappa(1-\kappa)} - \tfrac{\xi}{\kappa(\kappa+\xi)}\right) - \tfrac{1}{(\kappa+\xi)^2} = \frac{1}{\xi(1-\kappa)(\kappa+\xi)}.
$$

This is strictly positive on the domain, which implies strict concavity of $h(\kappa,\xi)$ and hence of $f(\kappa,\xi)$. The function $f(\kappa,\xi)$ therefore has a unique maximum. The value at the maximum is 0 by the asymptotic lower bounds of Theorem A.7 and the fact that $C(k,l) \le 1$. Let $\kappa_0$ and $\xi_0$ be the location of the maximum of $f(\kappa,\xi)$. Since $f(\kappa,\xi) = -\infty$ on the boundary of its domain, the maximum occurs in the interior. The concavity and differentiability properties imply that there exists $b_3 > 0$ such that if $((\kappa-\kappa_0)^2 + (\xi-\xi_0)^2)^{1/2} \ge \epsilon$, then $f(\kappa,\xi) \le -b_3\epsilon^2$ (this can be shown formally by use of the multidimensional Taylor series expansion with the remainder and applying strict concavity and boundedness of the domain). Choose $b_1$ small enough and $b_2$ large enough to compensate for the differences in the arguments of $r$ in $f_u(\kappa,\xi,N)$ and $f(\kappa,\xi)$. This gives $f_u(\kappa,\xi,N) \le -b_1\epsilon^2 + b_2\ln(N)/N$. ■

Lemma 3.7 allows us to consider only those $\binom{N}{k}\Delta_k(N,\alpha/N)$ with $k$

near the the maximizing value. Define

$$T_k(d,N) = \binom{N}{k} \mu_{\Delta(N)}(\mathbf{y} \in \Delta_k(N) \mid |\mathbf{1}/N - \mathbf{y}|_1 < d).$$

To estimate $T_k$, we will study its density function $T'_k(d,N) = \frac{d}{dt}T_k(t,N)|_{t=d}$. Note that $T_k(t,N)$ is differentiable.

**Lemma 3.8** *Let $\kappa_0$ be as in Lemma 3.7. There exist constants $b_i > 0$, $\delta_0 > 0$ and a function $0 < d(\kappa) < 2$ such that for $|\kappa - \kappa_0| \leq \delta_0$ and $|d - d(\kappa)| > \epsilon$, $T'_{\lfloor \kappa N \rfloor}(d,N) \leq e^{-b_1\epsilon^2 N + b_2 \ln(N)}$. The function $d(\kappa)$ can be chosen to be continuously differentiable on its domain.*

**Proof.** By using the first part of the proof of Lemma 3.7, we can write $T_k$ as follows:

$$\begin{aligned} T_k(d,N) &= \binom{N}{k}(N-1)!/N^{N-1} \\ &\quad \nu((\mathbf{y},\mathbf{z}) \mid \mathbf{y} \in \mathbf{R}^{k-1},\ \mathbf{y} \geq 0, \\ &\qquad \mathbf{z} \in [0,1]^{N-k},\ 2\mathbf{z}\cdot\mathbf{1} \leq Nd, \\ &\qquad \mathbf{y}\cdot\mathbf{1} \leq \mathbf{z}\cdot\mathbf{1}) \\ &= \binom{N}{k}(N-1)!/N^{N-1} \int_0^{Nd/2} dt \\ &\qquad U'(N-k,t)\frac{1}{(k-1)!}t^{k-1}. \end{aligned}$$

Differentiating by $d$ gives

$$\begin{aligned} T'_k(d,N) &= \binom{N}{k}(N-1)!/N^{N-2}\,(Nd/2)^{k-1}/(k-1)! \\ &\quad U'(N-k,Nd/2). \end{aligned}$$

Consider $k = \lfloor \kappa N \rfloor$ and define

$$t(d) = \lim_N \ln(T'_{\kappa N}(d,N))/N.$$

We proceed as in the proof of Lemma 3.7 and use Theorem A.7 to obtain:

$$\begin{aligned} t(d) &= H_e(\kappa) - 1 + (1-\kappa)h'(d/(2(1-\kappa))) \\ &\quad + \kappa\ln(d/2) + \kappa - \kappa\ln(\kappa), \end{aligned}$$

where $h'(x) = \lim_N \ln(U'(N,x))/N$ is strictly concave. It is clear that $t(d)$ is strictly concave in $d$, with a negative second derivative where it is finite. Hence $t$ has a unique maximum for some $d = d(\kappa)$, at which it must be 0. It follows that there is a constant $b_3 > 0$ such that for, $t(d(\kappa)-\epsilon) \leq e^{-b_3\epsilon^2 N}$. Since $h'(x) = -\infty$ for $x \geq 1$, and the second derivative of $\ln(d/2)$ is strictly bounded above by $c < 0$ for $d/2 \leq 1$, we can choose $b_3$ independently of $\kappa$. The derivative $\partial_d t$ is strictly monotone in $d$ for each $\kappa$ in the domain and is continuously differentiable in both $\kappa$ and $d$ (using Theorem A.7 for $h'$). Thus $d(\kappa)$ is defined by $\partial_d t(d(\kappa)) = 0$. By implicit differentiation, $\partial_d\partial_\kappa t \partial_\kappa d + \partial_\kappa^2 t = 0$. By strict concavity and continuity of the functions involved, $\partial_\kappa d$ is well defined with a continuous derivative.

To obtain the bound of the lemma, it now suffices to apply (5) of Theorem A.7 and Stirling's approximation. Note that $0 < \kappa_0 < 1$ so that the term $r'(x)$ in (5) is bounded for $\delta_0$ small enough. ■

We are now ready to give the proof of Theorem 3.5.

**Proof of Theorem 3.5.** The quantity $2/e$ is asymptotically the average distance of elements of $\Delta(N)$ to $\mathbf{1}/N$. Let $\kappa_0$ and $a_i$ be as in the statement of Lemma 3.7 and $d(\kappa)$, $b_i$ and $\delta_0$ as in the statement of Lemma 3.8. Let $d_0 = d(\kappa_0)$. Choose $c_3$ such that $|d_0 - d(\kappa_0 + t)| < c_3|t|$ for all $t < \delta_0$.

We claim that $d_0 = 2/e$. The results so far imply that the distribution of $|\mathbf{1}/N - \mathbf{y}|_1$ is strongly concentrated at its average as $N \to \infty$, which implies the result. More specifically, to see that $d_0 \leq 2/e + o(1)$, consider for $\delta/(2c_3) \leq \delta_0$

$$\begin{aligned} T(d_0 - \delta, N) &\leq \sum_{k:|k/N-\kappa_0|\leq\delta/(2c_3)} T_k(d_0-\delta, N) + \sum_{k:|k/N-\kappa_0|>\delta/(2c_3)} T_k(2,N) \\ &\leq e^{-b_1(\delta/2)^2N + b_2'\ln(N)} + e^{-a_1(\delta/(2c_3))^2N + a_2'\ln(N)}, \end{aligned}$$

where $a_2$ and $b_2$ have been adjusted to absorb factors of $N$ and 2 from the summation and integration of $T_k'$.

Let $d_a$ be the average value of $|\mathbf{1}/N - \mathbf{y}|_1$. The above inequalities imply that

$$d_a \geq (d_0 - \delta)(1 - e^{-a\delta^2 N(1+o(1))}).$$

A reverse inequality is obtained similarly and the claim follows by letting $N \to \infty$ and $\delta \to 0$.

Replacing $\delta$ by $\epsilon$ in the inequalities above and choosing $c_2$ large enough gives the theorem, provided that $\epsilon/(2c_3) \leq \delta_0$. One can extend the result to all $\epsilon$ by noting that only the case $\epsilon < 2$ is non-trivial and choosing $c_1$ small enough and $c_2$ large enough to cover the remaining range by exploiting monotonicity of $T(d_0 - \epsilon, N)$ for $\epsilon/(2c_3) > \delta_0$ and $N$ large enough. ■

### 3.3 On the Distribution of $|x - y| : y \in \Delta(N', N)$

Consider $\Delta(N', N)$ with $N' = o(N)$. We would like to show that for all $x \in \Delta(N)$, most elements of $\Delta(N', N)$ have distance at least $2 - \epsilon$.

**Theorem 3.9** *There exists constants $c_i > 0$ such that for $0 < \gamma < 1$*

$$\mu_{\Delta(\lfloor \gamma N \rfloor, N)}(\mathbf{y} \;\mid\; |\mathbf{x} - \mathbf{y}| < 2 - \epsilon) \leq e^{-(c_1 \epsilon - c_2 \gamma^{1/4})\gamma N + c_3 \ln(N)}.$$

**Proof.** We assume without loss of generality that $\lfloor \gamma N \rfloor = \gamma N$ (the correction to the exponent on the righthand side can be absorbed by the $c_3 \ln(N)$ term). Fix $N$ and let $\delta$ and $\rho$ be positive constants with properties to be determined. Let $\mathbf{x} \in \Delta(N)$. Define

$$\begin{aligned} L(\mathbf{x}) &= \{i \;\mid\; x_i \leq \delta/N\}, \\ B(S) &= \Delta(S, N) \cap \{\mathbf{y} \;\mid\; |\mathbf{x} - \mathbf{y}|_1 < 2 - \epsilon\} \end{aligned}$$

with $|S| = \gamma N$. Our goal is to show that for most $S$, $\mu_{\Delta(S,N)}(B(S))$ is small. To do so requires another lemma on the distribution of the TVD.

**Lemma 3.10** *Let $\mathbf{z} = (\mathbf{z}^{(1)}, \mathbf{z}^{(2)})$ with $\mathbf{z}^{(1)} \in \mathbf{R}^k$ and $\mathbf{z}^{(2)} \in \mathbf{R}^{N-k}$. Then for $k = \lfloor \kappa N \rfloor$,*

$$\mu_{\Delta(N)}(\mathbf{y} \;\mid\; |\mathbf{z} - \mathbf{y}| < |\mathbf{z}^{(1)} \cdot \mathbf{1}| + |1 - \mathbf{z}^{(2)} \cdot \mathbf{1}| - \epsilon) \leq e^{(\kappa |\ln(\kappa/e)| - \epsilon(1-\kappa)/2)N}.$$

**Proof.** For $\mathbf{y} \in \Delta(N)$, write $\mathbf{y} = (\mathbf{y}^{(1)}, \mathbf{y}^{(2)})$ with $\mathbf{y}^{(1)} \in \mathbf{R}^k$ and $\mathbf{y}^{(2)} \in \mathbf{R}^{N-k}$. Let $w_1 = \mathbf{y}^{(1)} \cdot \mathbf{1}$. We have

$$\begin{aligned} |\mathbf{z} - \mathbf{y}|_1 &\geq |\mathbf{z}^{(1)} \cdot \mathbf{1} - \mathbf{y}^{(1)} \cdot \mathbf{1}| + |\mathbf{z}^{(2)} \cdot \mathbf{1} - \mathbf{y}^{(2)} \cdot \mathbf{1}| \\ &= |\mathbf{z}^{(1)} \cdot \mathbf{1} - w_1| + |1 - w_1 - \mathbf{z}^{(2)} \cdot \mathbf{1}| \\ &\geq |\mathbf{z}^{(1)} \cdot \mathbf{1}| + |1 - \mathbf{z}^{(2)} \cdot \mathbf{1}| - 2w_1. \end{aligned}$$

It follows that

$$\mu_{\Delta(N)}(\mathbf{y} \;\mid\; |\mathbf{z} - \mathbf{y}|_1 < |\mathbf{z}^{(1)} \cdot \mathbf{1}| + |1 - \mathbf{z}^{(2)} \cdot \mathbf{1}| - \epsilon) \;\; \leq \;\; \mu(\mathbf{y} \;\mid\; w_1 > \epsilon/2).$$

The distribution of $w_1$ for $\mathbf{y}$ in $\Delta(N)$ is that of a $\beta$ distribution:

$$f(w_1) = (N-1)\binom{N-2}{k-1} w_1^{k-1}(1-w_1)^{N-k-1}.$$

We can estimate

$$\begin{aligned}
\mu(\mathbf{y} \mid w_1 > \epsilon/2) &\leq \int_{\epsilon/2}^{1} dt (N-1)\binom{N-2}{k-1}(1-t)^{N-k-1} \\
&= \binom{N-1}{k-1}(1-\epsilon/2)^{N-k} \\
&\leq \binom{N}{k}(1-\epsilon/2)^{N-k} \\
&\leq e^{\kappa|\ln(\kappa/e)|N + \ln(1-\epsilon/2)(1-\kappa)N} \\
&\leq e^{\kappa|\ln(\kappa/e)|N - \epsilon(1-\kappa)N/2},
\end{aligned}$$

where we used Lemma A.3 and its corollary. ■

Suppose that $|S \cap L(\mathbf{x})| = (1-\xi)|S| = (1-\xi)\gamma N$. We can estimate $\mu_{\Delta(S,N)}(B(S))$ with the help of Lemma 3.10 by projecting $\mathbf{x}$ on the coordinates in $S$ and considering the coordinates in $S \setminus L(x)$ versus those in $S \cap L(\mathbf{x})$ . Let $w_1$, $w_2$ and $w_3$ be the total weight of the coordinates of $\mathbf{x}$ in $S \setminus L(\mathbf{x})$, $S \cap L(\mathbf{x})$ and the complement of $S$, respectively. We have $w_2 \leq (1-\xi)\delta\gamma$ and $w_1 + w_3 \geq 1 - (1-\xi)\delta\gamma$. The distance parameter in Lemma 3.10 relative to $S$ partitioned into $S \setminus L(x)$ and $S \cap L(x)$ is given by $w_1 + 1 - w_2$. The distance of $\mathbf{x}$ to an element of $D(S,N)$ due to the coordinates outside of $S$ is $w_3$. We have $w_1 + 1 - w_2 + w_3 \geq 2 - \epsilon/2$, provided that $\delta\gamma \leq \epsilon/4$. Write $a(\xi) = -\xi|\ln(\xi/e)|) + \epsilon(1-\xi)/4$. Lemma 3.10 implies that

$$\mu_{\Delta(S,N)}(B(S)) \quad \leq \quad e^{-a(\xi)\gamma N}.$$

Let $a = a(\rho)$ with $0 < \rho < 1$. Since $a(\xi)$ is decreasing in $\xi$, we have $\mu_{\Delta(S,N)}(B(S)) \leq e^{-a\gamma N}$ provided that $|S \cap L((x))| \geq |S| - \lceil \rho |S| \rceil$ and $\delta\gamma \leq \epsilon/4$.

We estimate the fraction of subsets $S$ satisfying $|S \cap L(x)| < |S| - \lceil \rho |S| \rceil$. Note that $(N - |L(x)|)\delta/N \leq 1$, so that $|L(x)|/N \geq 1 - 1/\delta$. If $1/\delta \leq \rho$ we can apply Lemma A.6 and monotonicity of $K_e$ to obtain

$$\begin{aligned}
&\left|\{S \mid |S| = \gamma N,\ |S \cap L(x)| \leq |S| - \lceil \rho |S| \rceil\}\right| \\
&\leq \gamma N \binom{N - \lfloor N/\delta \rfloor}{N - \lceil \rho\gamma N \rceil}\binom{\lfloor N/\delta \rfloor}{\lceil \rho\gamma N \rceil}
\end{aligned}$$

$$
\begin{aligned}
&\leq \binom{N}{\gamma N} e^{-K_e(\rho,1/\delta)\gamma N+\ln(\gamma N)} \\
&\leq e^{(-\rho\ln(\rho\delta)-1/\delta+\rho)\gamma N+\ln(\gamma N)} \\
&\leq e^{-\ln(\rho\delta/e)\rho\gamma N+\ln(\gamma N)}.
\end{aligned}
$$

Combining these results we get

$$
\begin{aligned}
\mu_{\Delta(\gamma N,N)}(\mathbf{y} \mid |\mathbf{x}-\mathbf{y}| < 2-\epsilon) \quad \leq \quad & e^{-(\epsilon(1-\rho)/4-\rho|\ln(\rho/e)|)\gamma N} \\
& + e^{-\ln(\delta\rho/e))\rho\gamma N+\ln(\gamma N)},
\end{aligned}
$$

for $0 < \rho < 1$, $\delta\rho \geq 1$ and $\delta\gamma \leq \epsilon/4$.

A rough estimate can be obtained by letting $\rho = \epsilon/(16|ln(\epsilon/(16e))|)$. and $\delta = \epsilon/(4\gamma)$. We will assume that $\gamma \leq (\epsilon/(16e^2))\epsilon^2/(64|ln(\epsilon/(16e))|)$. This is true for $\gamma \leq c_2'\epsilon^4$ for some constant $c_2'$. Recall that without loss of generality $\epsilon < 2$. Thus $\ln(8e) \leq |ln(\epsilon/(16e))| \leq 16e/\epsilon$.

$$
\begin{aligned}
\epsilon(1-\rho)/4 &\geq 15\epsilon/64, \\
|ln(\rho/e)| &= |ln(\epsilon/(16e)) - \ln|ln(\epsilon/(16e))|| \\
&\leq 2|ln(\epsilon/(16e))| \\
\rho|ln(\rho/e)| &\leq \epsilon/8 \\
\ln(\rho\delta/e) &\geq |\ln(\epsilon/(16e)))| \\
\rho\ln(\rho\delta/e) &\geq \epsilon/16.
\end{aligned}
$$

The inequality of the theorem follows. ■

### 3.4 Extensions of the bounds to $\Delta(N', N)^k$

It is now straightforward to obtain general bounds for $\Delta(N', N)^k$ by using Lemma A.1.

**Theorem 3.11** *There exist $c_i > 0$ such that for $1 \leq k \leq N$ and $\mathbf{x} \in \Delta(N)^k$*

$$
\mu_{\Delta(N)^k}(\mathbf{y} \mid |\mathbf{x}-\mathbf{y}|_1 < 2/e - \epsilon) \leq e^{-(c_1\epsilon^2 N + c_2\ln(N))k}.
$$

**Proof.** Theorem 3.5 and Lemma A.1 with $m = 2k$ give

$$
\begin{aligned}
\mu_{\Delta(N)^k}(\mathbf{y} \mid |\mathbf{x}-\mathbf{y}|_1 < d) \quad &\leq \binom{2k-1}{k-1} (e^{-c_1\epsilon^2/4\, N + c_2\ln(N)})^k \\
&\leq e^{-c_1'\epsilon^2 Nk + c_2'\ln(N)k},
\end{aligned}
$$

for suitable choices of constants. ∎

**Theorem 3.12** *There exist $c_i > 0$ such that for $0 < \gamma < 1$*

$$\mu_{\Delta(\gamma N,N)^k}(\mathbf{y} \mid |\mathbf{x}-\mathbf{y}| < 2-\epsilon) \leq e^{-(c_1\epsilon - c_2\gamma^{1/4})\gamma N k + c_3 \ln(N) k}.$$

**Proof.** Follow the proof of Theorem 3.11, using Theorem 3.9 and Lemma A.1 with $m = 2k$. ∎

# A Appendix

## A.1 Miscellaneous Bounds

We begin by giving several lemmas which are special cases of weak large deviation laws.

**Lemma A.1** *Let $\mu_i$ be probability distributions on $\mathbf{R}$ and $\mu^{(n)} = \prod_{i=1}^n \mu_i$. Suppose that $\mu_i(x \mid x > t) \leq e^{-\lambda(t)}$, with $\lambda(t)$ convex (where finite) and for $t \leq 0$, $\lambda(t) = 0$. Then for $m > n$,*

$$\mu^{(n)}(\mathbf{x} \mid \mathbf{x}\cdot\mathbf{1} \geq nt) \quad \leq \quad \binom{m-1}{n-1} e^{-n\lambda(t(1-n/m))}.$$

**Proof.** Let $m > n$. Consider $\mathbf{x} \in \mathbf{R}^n$ such that $\mathbf{x}\cdot\mathbf{1} \geq nt$. If $y$ is the vector with coordinates $y_i = \lfloor x_i m/nt \rfloor nt/m$, then $\mathbf{y}\cdot\mathbf{1} \geq nt(1-n/m)$. It follows that for each such $\mathbf{x}$, there is an integer vector $\mathbf{l}$ such that $\mathbf{l}\cdot\mathbf{1} = m-n$ and $\mathbf{l}nt/m \leq \mathbf{x}$. Define $\sigma(l_i) = l_i$ for $l_i > 0$ and $\sigma(l_i) = -\infty$ otherwise. Using the assumption that $\lambda(0) = 0$, we can estimate

$$\begin{aligned}
\mu^{(n)}(\mathbf{x} \mid \mathbf{x}\cdot\mathbf{1} \geq nt) \quad &\leq \sum_{\mathbf{l}:\mathbf{l}\in\mathbf{Z}^n,\, \mathbf{l}\geq 0,\, \mathbf{l}\cdot\mathbf{1}=m-n} \mu_i(\mathbf{x} \mid \mathbf{x} \geq \sigma(\mathbf{l})) \\
&\leq \sum_{\mathbf{l}:\mathbf{l}\in\mathbf{Z}^n,\, \mathbf{l}\geq 0,\mathbf{l}\cdot\mathbf{1}=m-n} e^{-\sum_{i=1}^n \lambda(l_i nt/m)}.
\end{aligned}$$

Convexity of $\lambda$ implies that

$$\sum_{i=1}^n \lambda(l_i nt/m) \quad \leq \quad n\lambda(\sum_i l_i t/m).$$

This gives

$$\mu^{(n)}(x \mid x \cdot \mathbf{1} \geq nt) \leq \sum_{l : l \in \mathbf{Z}^n,\, l \geq 0, l \cdot \mathbf{1} = m-n} e^{-n\lambda(t(1-n/m))}$$

$$\leq \binom{m-1}{n-1} e^{-n\lambda(t(1-n/m))}.$$

■

Let $H_e(\kappa) = -\kappa \ln(\kappa) - (1-\kappa)\ln(1-\kappa)$. This is the information function base $e$.

**Lemma A.2** *For* $0 \leq \kappa \leq 1$, $H_e(\kappa) \leq \kappa |\ln(\kappa/e)|$.

**Proof.** The summand $-(1-\kappa)\ln(1-\kappa)$ is concave with a slope of 1 at $\kappa = 0$. ■

**Lemma A.3** *For* $n \geq 1$ *and* $0 \leq \kappa \leq 1$,

$$\binom{n}{\lfloor \kappa n \rfloor} \leq e^{H_e(\lfloor \kappa n \rfloor / n) n}$$

$$\leq e^{H_e(\kappa) n + \ln(en)}$$

*and* $\lim_n \ln \binom{N}{\lfloor \kappa n \rfloor} / n = H_e(\kappa)$.

**Proof.** For $\kappa n$ integral, it can be shown that $\binom{n}{\kappa n} \leq e^{H_e(\kappa) n}$ by applying a tight form of Stirling's approximation, for example,

$$\sqrt{2\pi n}(n/e)^n e^{1/(12n+1)} \leq n! \leq \sqrt{2\pi n}(n/e)^n e^{1/(12n)}.$$

This form of Stirling's approximation can be found in [6]. For non-integral $\kappa n$ it suffices to observe that $|H_e(\kappa) - H_e(\lfloor \kappa n \rfloor / n)| \leq H_e(\frac{1}{n})$. The result then follows by Lemma A.2. ■

**Corollary A.4** *For* $0 \leq \kappa \leq 1$, $\binom{n}{\lfloor \kappa n \rfloor} \leq e^{\kappa |\ln(\kappa/e)|) n}$.

**Proof.** Let $\kappa' = \lfloor \kappa n \rfloor / n$. By Lemma A.3 we have

$$\begin{aligned} \binom{n}{\kappa' n} &\leq e^{H_e(\kappa')n} \\ &\leq e^{\kappa' |\ln(\kappa'/e)| n} \\ &\leq e^{\kappa |\ln(\kappa/e)| n}. \end{aligned}$$

■

For $0 < \kappa < 1$ and $0 \leq \xi \leq 1$, define $K_e(\xi, \kappa) = \xi \ln(\xi/\kappa) + (1-\xi)\ln((1-\xi)/(1-\kappa))$. Also let $K_e(0,0) = K_e(1,1) = 0$ and $K_e(\xi, 0) = K_e(\xi, 1) = \infty$ otherwise.

**Lemma A.5** *For* $0 < \kappa \leq \xi$, $K_e(\xi, \kappa) \geq \xi \ln(\xi/\kappa) + \kappa - \xi$.

**Proof.**

$$\begin{aligned} K_e(\xi, \kappa) &= \xi \ln(\xi/\kappa) + (1 - xi)\ln((1-\xi)/(1-\kappa)) \\ &\geq \xi \ln(\xi/\kappa) + (1-\xi)(1 - (1-\kappa)/(1-\xi)) \\ &= \xi \ln(\xi/\kappa) + \kappa - \xi, \end{aligned}$$

since $\ln(x) \geq (1 - 1/x)$ for $0 < x \leq 1$. ■

**Lemma A.6** *Let* $0 \leq \gamma \leq 1$ *and* $\gamma' = \lfloor \gamma n \rfloor / n$. *For* $0 \leq \kappa \leq \xi \leq 1$ *and* $1 \leq n$,

$$\binom{n - \lfloor \kappa n \rfloor}{\gamma' n - \lceil \xi \gamma' n \rceil} \binom{\lfloor \kappa n \rfloor}{\lceil \xi \gamma' n \rceil} \leq \binom{n}{\gamma' n} e^{-K_e(\xi,\kappa)\gamma' n}.$$

**Proof.** Define $(n)_l = n(n-1)\ldots(n-l+1)$ (the $l$'th falling factorial of $n$). Assume first that $\kappa n$ and $\xi\gamma' n$ are integral, and ignore the restriction that $\kappa \leq \xi$. The inequality is trivial for $\xi\gamma' > \kappa$.

$$\begin{aligned} \binom{(1-\kappa)n}{(1-\xi)\gamma' n}\binom{\kappa n}{\xi \gamma' n} &= \binom{n}{\gamma' n}\binom{\gamma n}{\xi \gamma' n} ((1-\kappa)n)_{(1-\xi)\gamma' n} (\kappa n)_{\xi \gamma' n} \,/\, (n)_{\gamma' n} \\ &\leq \binom{n}{\gamma' n}\binom{\gamma' n}{\xi \gamma' n} ((1-\kappa)n)^{(1-\xi)\gamma' n} (\kappa n)^{\xi \gamma' n} \,/\, n^{\gamma' n} \\ &\leq \binom{n}{\gamma' n} e^{-K_e(\xi,\kappa)\gamma' n}, \end{aligned}$$

where we applied the inequality of Corollary A.4 and estimated the term involving the falling factorials by using the inequality $(a-c)/(b-c) \leq a/b$ for $b \geq a$ and $a > c \geq 0$.

If $\kappa n$ and $\xi\gamma n$ are not integral, the inequality holds with $\kappa$ and $\xi$ replaced by $\kappa' = \lfloor \kappa n \rfloor / n$ and $\xi' = \lceil \xi\gamma' n \rceil$. The result follows because the exponent on the righthand side of the desired inequality is increasing in $\kappa$ and decreasing in $\xi$ for $\kappa \leq \xi$. ∎

## A.2 Cramér's Theorem for the Uniform Distribution

One of the fundamental results of the theory of large deviations is Cramér's theorem. Here we need a version of this theorem for uniformly distributed random variables.

**Theorem A.7** *Let $X_i$ be independent and uniformly distributed on $[-1,1]$ and write $S_n = \frac{1}{n}\sum_{i=1}^n X_i$. Define $F(x) = \text{Prob}(S_n < x)$ and let $f(x) = F'(x)$ be the density of $S_n$. Let $r(x) = -\lim_n \ln(f(x))/n$. Then the following hold:*

(1) *$r(x) \geq 0$, $r(0) = 0$ and $r(x) = \infty$ for $x \notin (-1,1)$.*

(2) *$r(x)$ is convex and twice differentiable on $(-1,1)$.*

(3) *For $x \leq 0$, $F(x) \leq e^{-r(x)n}$.*

(4) *For $x \leq 0$, $r(x) = -\lim_n \ln(F(x))/n$.*

(5) *There exists $c$ such that for $-1 < x < 1$ $f(x) \leq e^{-r(x)n + c\ln((|r'(x)|+e)n)}$.*

(6) *$r(x) = -\lim_n \ln(f(x))/n$.*

**Proof.** The function $r$ is the *rate function*. In this case it is obtained as follows. Let

$$\Gamma(t) = \ln \text{Exp}(e^{tX_i}) = \ln(\frac{2}{t}\sinh(t)).$$

The function $r(x)$ is given by $r(x) = \sup_t(tx - \Gamma(t))$. Since $\Gamma(t)$ is smooth and strictly convex, $r(x)$ is obtained by first finding $t(x)$ such that $\Gamma'(t(x)) = x$ and then evaluating $r(x) = t(x)x - \Gamma(t(x))$. By implicit differentiation, $\Gamma''(t(x))t'(x) = 1$. By strict convexity, $\Gamma''(t)$ is never zero, so $t$ is continuously differentiable on its domain. By taking higher derivatives implicitly, one can see that $t$ is in fact smooth (where finite). This implies that $r$ is smooth where finite. Note that $r'(x) = t(x)$. This together with the proof of Cramér's theorem found in most textbooks gives (1), (2), (3) and (4) (e.g. [3]). For the inequality of (5) observe that $f(x)$ is symmetric and unimodular so that for $x \leq 0$ and $\delta > 0$, $F(x) \geq \delta f(x-\delta)$. Hence for $\delta \leq |x|$,

$$\begin{aligned} f(x) &= f(x+\delta-\delta) \\ &\leq \frac{1}{\delta} F(x+\delta) \\ &\leq \frac{1}{\delta} e^{-r(x+\delta)n}. \end{aligned}$$

If $|x| \geq \frac{1}{|r'(x)|n}$, let $\delta = \frac{1}{|r'(x)|n}$ and use convexity of $r$ to see that $f(x) \leq |r'(x)|ne^{-r(x)n+1} \leq e^{-r(x)n+\ln((|r'(x)|+e)n)}$. For $|x| \leq \frac{1}{|r'(x)|n}$ we use the result on cube slicing in [1] which implies that $f(0) \leq \sqrt{n/2}$. For such $x$ we have $r(x) \leq \frac{1}{n}$. Hence $f(x) \leq \sqrt{n/2} \leq e^{-r(x)n+\ln((|r'(x)|+e)n)}$. For $x = 0$, (5) is trivial, and for $x > 0$ we can use symmetry. Part (6) follows from (4), (5) and the observation that for $x \leq 0$, $F(x) \leq (1-x)f(x)$. ■

**Acknowledgements.** Thanks to the enthusiastic support of the discrete mathematics group at the Los Alamos National Laboratory. Special thanks to Vance Faber for discussions of the geometry of $\Delta(N)$ and the hypercube.

## References

[1] K. Ball. Cube slicing in $\mathbf{R}^n$. *Proceedings of the AMS*, 97:465–473, 1986.

[2] A. Barenco, C. H. Bennett, R. Cleve, D. P. DiVincenzo, N. Margolus, P. Shor, and T. Sleator. Elementary gates for quantum computation. Submitted to Physical Review A, 1995.

[3] A. Dembo and O. Zeitouni. *Large Deviation Techniques and Applications.* Jones and Bartlett Publishers, 1993.

[4] D. P. DiVincenzo. Two-bit gates are universal for quantum computation. *Phys. Rev. A*, 51:1015–1022, 1995.

[5] E. Knill. Approximation by quantum circuits. Technical Report LAUR-95-2225 and `68Q-95-29 at http://www.c3.lanl.gov/laces`, Los Alamos National Laboratory, 1995.

[6] H. Robbins. A remark on Stirling's formula. *American Mathematical Monthly*, 62:26–29, 1955.

[7] D. R. Simon. On the power of quantum computation. In *Proceedings of the 35'th Annual Symposium on Foundations of Computer Science*, pages 116–123. IEEE Press, 1994.

[8] A. Yao. Quantum circuit complexity. In *Proceedings of the 34th Annual Symposium on Foundations of Computer Science*, pages 352–360. IEEE Press, 1993.